\documentclass[prl,twocolumn,showkeys,showpacs,superscriptaddress,reprint]{revtex4}
\usepackage{graphicx,latexsym,amssymb,amsmath}

\begin{document}

\title{Herschel-Bulkley Shearing Rheology Near the Athermal Jamming Transition}

\author{Peter Olsson}
\affiliation{Department of Physics, Ume{\aa} University, 901 87 Ume{\aa}, Sweden}
\author{S. Teitel}
\affiliation{Department of Physics and Astronomy, University of
Rochester, Rochester, NY 14627}
\date{\today}

\begin{abstract}
We consider the rheology of soft-core frictionless disks in two dimensions in the neighborhood of the athermal jamming transition.  From numerical simulations of bidisperse, overdamped, particles, we argue that the divergence of the viscosity below jamming is characteristic of the hard-core limit, independent of the particular soft-core interaction.  We develop a mapping from soft-core to hard-core particles that recovers all the critical behavior found in earlier scaling analyses.  Using this mapping we derive a  relation that gives the exponent of the non-linear Herschel-Bulkley rheology above jamming in terms of the exponent of the diverging viscosity below jamming.
\end{abstract}
\pacs{45.70.-n, 64.60.-i, 83.80.Fg}
\maketitle

A variety of disordered soft solids, such as foams, colloids and emulsions, are empirically observed to obey a non-linear rheology under steady state shearing known as the Herschel-Bulkley (HB) law \cite{HerschelBulkley}, $\sigma = \sigma_0 + c\dot\gamma^b$. Here $\dot\gamma$ is the shear strain rate, $\sigma$ is the average shear stress, $\sigma_0$ is the yield stress at $\dot\gamma\to 0$, and $b$ is an exponent usually (but not always) found experimentally to be in the range $0.33$ to $0.5$ \cite{SchallHecke}.  Detailed microscopic models of the viscous interaction in foams and emulsions have been studied to try and understand the HB form \cite{Denkov}.  However in the limit of very slow strain rates, $\dot\gamma\to 0$, it seems likely that the rheology will be determined by collective effects and may be characterized as a critical phenomenon \cite{Sollich,Caroli,Martens}.  In this limit, the HB rheology has been treated in terms of a phenomenological model of slow glassy relaxation \cite{Sollich}, and more recently in terms of the nucleation of localized plastic events \cite{Caroli}.  

Here we investigate this problem using numerical simulations of a model of athermally sheared, frictionless, soft-core disks.  
Such systems display a sharp {\it jamming transition} as the packing fraction $\phi$ increases.  For $\phi <\phi_J$, the system is liquid-like: at sufficiently small $\dot\gamma$ the rheology is linear  \cite{OlssonTeitelPRL} with a finite shear viscosity $\eta\equiv\sigma/\dot\gamma$ that diverges as $\eta\sim|\phi-\phi_J|^{-\beta}$ as $\phi\to\phi_J$.  For $\phi>\phi_J$ 
rheology is non-linear with a finite yield stress $\sigma_0$.  By numerically establishing a mapping from sheared soft-core particles to sheared hard-core particles, we propose a  relation between the exponent $b$ of the non-linear HB rheology above $\phi_J$ and the exponent $\beta$ of the diverging viscosity of the linear rheology below $\phi_J$.  

Our model \cite{OHern} is one of $N$ bidisperse soft-core disks in two dimensions (2D), with equal numbers of particles with radii ratio 1.4.  The soft-core interaction between two overlapping particles $i$ and $j$ is $V(r_{ij})=(\epsilon/\alpha)\delta_{ij}^\alpha$, where $\delta_{ij}=(1-r_{ij}/d_{ij})$ is the relative particle overlap. Here $r_{ij}$ is the particles center to center distance, $d_{ij}$ is the sum of their radii, and $\alpha=2$ or $5/2$ for harmonic or Hertzian interaction, respectively.  Lengths are measured in units of the small particle diameter $d_s$, and energy in units of $\epsilon$.  We use Durian's ``mean-field" dynamics \cite{Durian} of overdamped particles with a viscous dissipation with respect to the imposed average linear shear velocity flow,
\begin{equation}
\frac{d{\bf r}_i}{dt} =-C\sum_j\frac{dV(r_{ij})}{d{\bf r}_i}+y_i\dot\gamma\hat x\enspace.
\label{eDurian}
\end{equation}
Time is measured in units of $d_s/C\epsilon$.  Lees-Edwards boundary conditions \cite{LeesEdwards} induce a uniform shear strain rate $\dot\gamma$.  We use $N= 65536$ particles so that finite size effects are negligible.  Simulating at fixed $\dot\gamma$ and $\phi$, we compute the steady state time average of the elastic part of the pressure tensor \cite{OHern}, to define the scalar pressure $p$ and shear stress $\sigma$.  
We consider the pressure analog of the shear viscosity $\eta_p\equiv p/\dot\gamma$, rather than $\eta$, and restrict our analysis to a very narrow range about $\phi_J$, specifically $0.835\le\phi\le 0.846$ and $\dot\gamma\le 10^{-6}$ for harmonic, and $\dot\gamma\le 10^{-7}$ for Hertzian, so as to allow us to ignore effects due to corrections to scaling \cite{OlssonTeitelPRE,footnote}.

\begin{figure}[h!]
\begin{center}
\includegraphics[width=3.5in]{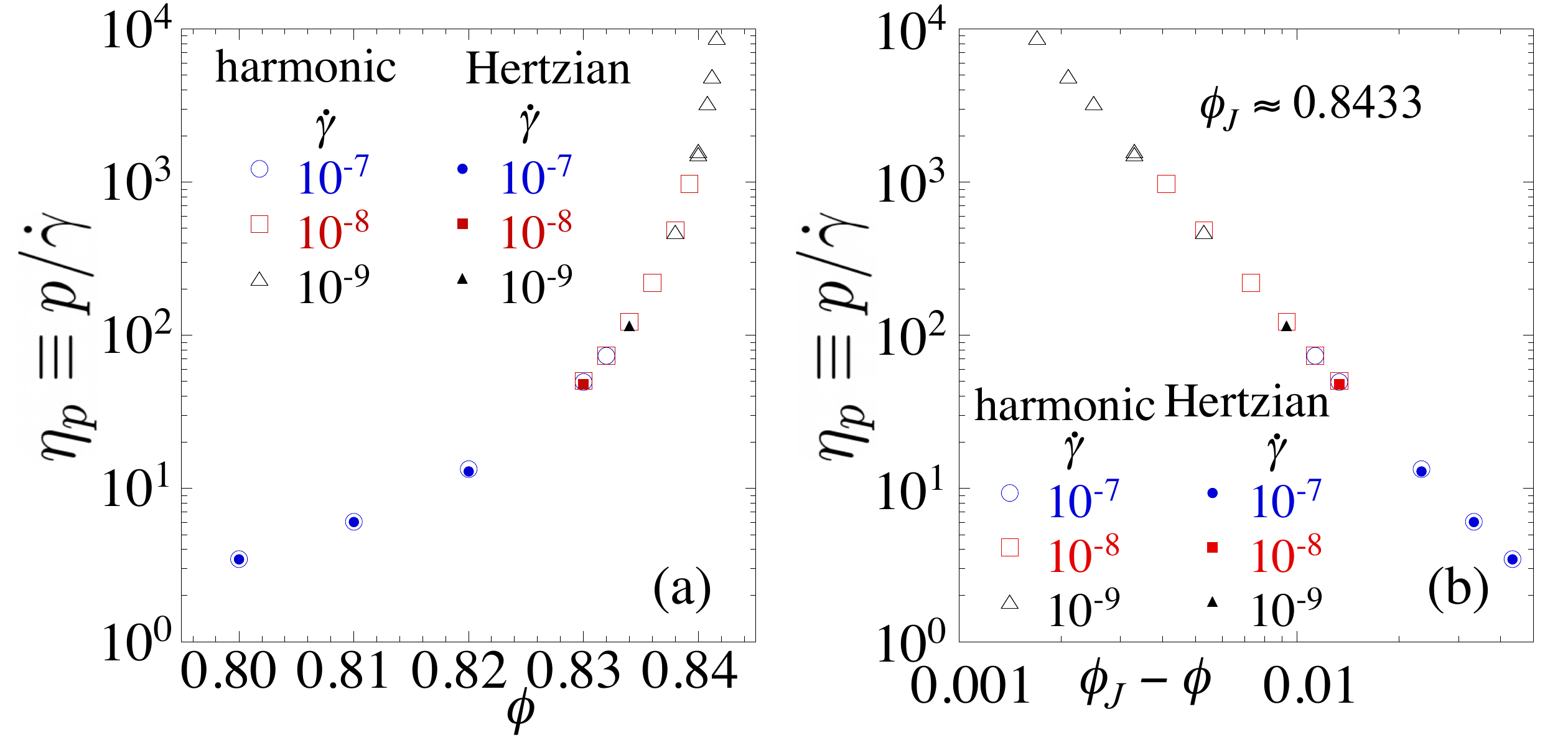}
\caption{(color online) (a) $\eta_p\equiv p/\dot\gamma$ vs $\phi$ for data below $\phi_J$ in the linear rheology region. Open symbols are for the harmonic interaction, solid symbols are for Hertzian. (b) Same data as in (a) but plotted vs $\phi_J-\phi$ with $\phi_J= 0.8433$.  
}
\label{f1}
\end{center}
\end{figure}

First we demonstrate the existence of the hard core limit below $\phi_J$.
In Fig.~\ref{f1} we compare $\eta_p$ for both harmonic and Hertzian interations, for small $\dot\gamma$ in the linear rheology region.
We see excellent agreement, showing that the $\dot\gamma\to 0$ limit of $\eta_p$ is independent of the particular soft-core interaction.  
For the strict hard-core limit, one expects that particles at different strain rates $\dot\gamma$ follow the same path through phase space, only with different velocities, ${\bf v}_i\propto\dot\gamma$.
For {\it overdamped} particles this implies that the contact forces also obey $f_{ij}\propto\dot\gamma$, and hence $p\propto\dot\gamma$.  One may think of $p/\dot\gamma$ in athermal shear driven flow as analogous to the virial $p/T$ of equilibrium hard-core particles.  In Fig.~\ref{f1}b we replot $\eta_p$ vs $\phi_J-\phi$, using $\phi_J=0.8433$ as previously determined \cite{OlssonTeitelPRE,Heussinger}.  We see a clear algebraic divergence  of $\eta_p$ over four decades as $\phi\to\phi_J$, demonstrating that the exponent $\beta$ is characteristic of the hard-core limit, independent of the particular soft-core interaction.  

We next consider behavior outside the linear rheology (hard core) region, showing that one can map soft-core particles at a given $\phi$ and $\dot\gamma$ onto an equivalent hard-core system at a lower $\phi_{\rm eff}(\phi,\dot\gamma)$, i.e. $\eta_p(\phi,\dot\gamma)=\eta_p(\phi_{\rm eff},\dot\gamma\to 0)\equiv\eta_p^{\rm hd}(\phi_{\rm eff})$, using a simple form for $\phi_{\rm eff}$.  If this mapping holds, then even outside the linear rheology region $\eta_p$ will have the scaling form,
\begin{equation}
\eta_p(\phi,\dot\gamma)=\eta_p^{\rm hd}(\phi_{\rm eff})=A(\phi_J-\phi_{\rm eff})^{-\beta}
\label{e-etap}
\end{equation}
with $\eta_p^{\rm hd}(\phi)$ given by the curve in Fig.~\ref{f1}.  Such a  mapping was suggested by early work  \cite{BarkerHenderson,AWC}  in equilibrium.  More recently Berthier and Witten \cite{BW} combined such a $\phi_{\rm eff}$ approach with critical scaling to study the equilibrium glass transition of soft spheres. A related analysis was done by Xu et al. \cite{Xu}, while
more recent works have sought to extend this mapping over wider ranges of pressure \cite{Schmiedeberg} and to systems with applied uniform shear strain rate $\dot\gamma$ \cite{Haxton}, though still at finite $T$.

Here we apply these ideas to an athermal shear-driven system. We follow Berthier and Witten \cite{BW} and assume that $\phi_{\rm eff}$ is set by the extent of particle overlaps, as measured by the average interaction energy per particle $E$.  We thus make the Ansatz,
\begin{equation}
\phi_{\rm eff}(\phi,\dot\gamma)=\phi-h(E(\phi,\dot\gamma))
\label{e-phieff0}
\end{equation}
with $h(E)$ an appropriate function to be determined.  Since $E=0$ when there are no overlaps, $h(0)=0$.

We can now determine $h(E)$ asymptotically close to $\phi_J$ by applying two simple conditions on $\phi_{\rm eff}$:
\begin{equation}
\phi_{\rm eff}(\phi,\dot\gamma\to 0) = \phi\enspace,\quad{\rm for}\>\phi<\phi_J\enspace.
\label{e-phieff1}
\end{equation}
Since $E\to 0$ as $\dot\gamma\to 0$ below $\phi_J$, overlaps vanish and $\phi_{\rm eff}=\phi$. 
The second condition is,
\begin{equation}
\phi_{\rm eff}(\phi,\dot\gamma\to 0) = \phi_J\enspace,\quad{\rm for}\>\phi>\phi_J\enspace.
\label{e-phieff2}
\end{equation}
Since $p\to p_0>0$ as $\dot\gamma\to 0$ for all $\phi>\phi_J$, then $\eta_p\to\infty$ everywhere along the dynamic yield stress curve. In a hard-core system, $\eta_p\to\infty$ only at $\phi=\phi_J$ ($\phi>\phi_J$ being excluded by the non-overlapping condition).
Thus, as $\dot\gamma\to 0$, all $\phi>\phi_J$ in a soft-core system must map onto $\phi_{\rm eff}=\phi_J$ of the equivalent hard-core system.

If we similarly define $E_0(\phi)\equiv E(\phi,\dot\gamma\to 0)$, then Eqs.~(\ref{e-phieff0}) and (\ref{e-phieff2}) imply $h(E_0(\phi))=\phi-\phi_J$ for $\phi>\phi_J$.  Close to $\phi_J$, $E_0$ scales to zero algebraically, $E_0\sim (\phi-\phi_J)^{y_E}$.  We thus conclude that $h(E)\sim E^{1/y_E}$, and so,
\begin{equation}
\phi_{\rm eff}(\phi,\dot\gamma) =\phi -c [E(\phi,\dot\gamma)]^{1/y_E}\enspace .
\label{e-phieff4}
\end{equation}

\begin{figure}[h!]
\begin{center}
\includegraphics[width=3.5in]{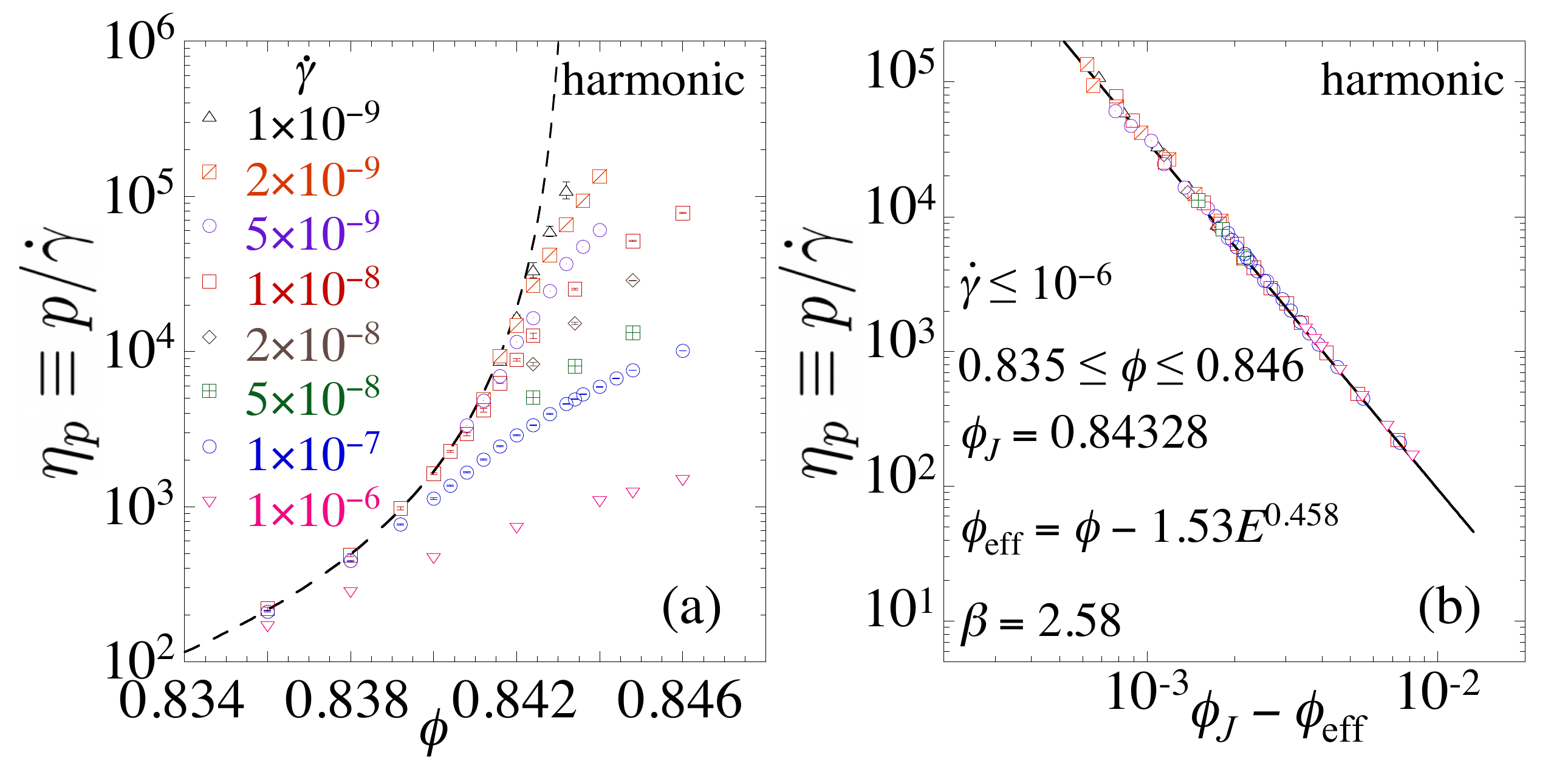}
\caption{(color online) (a) $\eta_p\equiv p/\dot\gamma$ vs $\phi$ for data above and below $\phi_J$ for harmonic soft core particles. (b) Same data as in (a) but plotted vs $\phi_J-\phi_{\rm eff}$. Dashed line in (a) and solid line in (b) is the fit to the model of Eqs.~(\ref{e-etap}) and (\ref{e-phieff4}). Symbols in (b) correspond to the legend given in (a).
}
\label{f2}
\end{center}
\end{figure}

We test this mapping by measuring $\eta_p$ and $E$ at various $\phi$ and $\dot\gamma$, and fitting our data to Eqs.~(\ref{e-etap}) and (\ref{e-phieff4}), taking $A$, $\phi_J$, $\beta$, $c$ and $y_E$ as free fitting parameters.  In Fig.~\ref{f2}a we show our raw data $\eta_p$ vs $\phi$  for the harmonic interaction,  including points both {\it above} and {\it below} $\phi_J$; $\eta_p$ decreases with increasing $\dot\gamma$, showing that our data include points outside the linear rheology region.  In Fig.~\ref{f2}b we show the results of our fit to the $\phi_{\rm eff}$ model, plotting $\eta_p$ vs $\phi_J-\phi_{\rm eff}$.  We find an excellent data collapse, yielding $\phi_J=0.84328\pm 0.00007$, $\beta=2.58\pm0.10$ and $y_E=2.18\pm0.02$ \cite{footnote}.  We can compare this to our earlier results \cite{OlssonTeitelPRE} from a more general critical scaling analysis that was independent of any $\phi_{\rm eff}$ assumption. Defining the pressure exponent $y_p$ by $p_0(\phi)\sim (\phi-\phi_J)^{y_p}$, our earlier results gave,
 $\phi_J=0.84347\pm 0.00020$, $\beta=\nu z-y_p=2.77\pm 0.20$ \cite{OTnote}; taking $y_E=2y_p$ for harmonic interaction, we get $y_E=2.16\pm0.06$.  The excellent agreement between the two approaches establishes the validity of our soft to hard-core mapping, $\phi_{\rm eff}$, for the range of data we simulate.

\begin{figure}[h!]
\begin{center}
\includegraphics[width=3.5in]{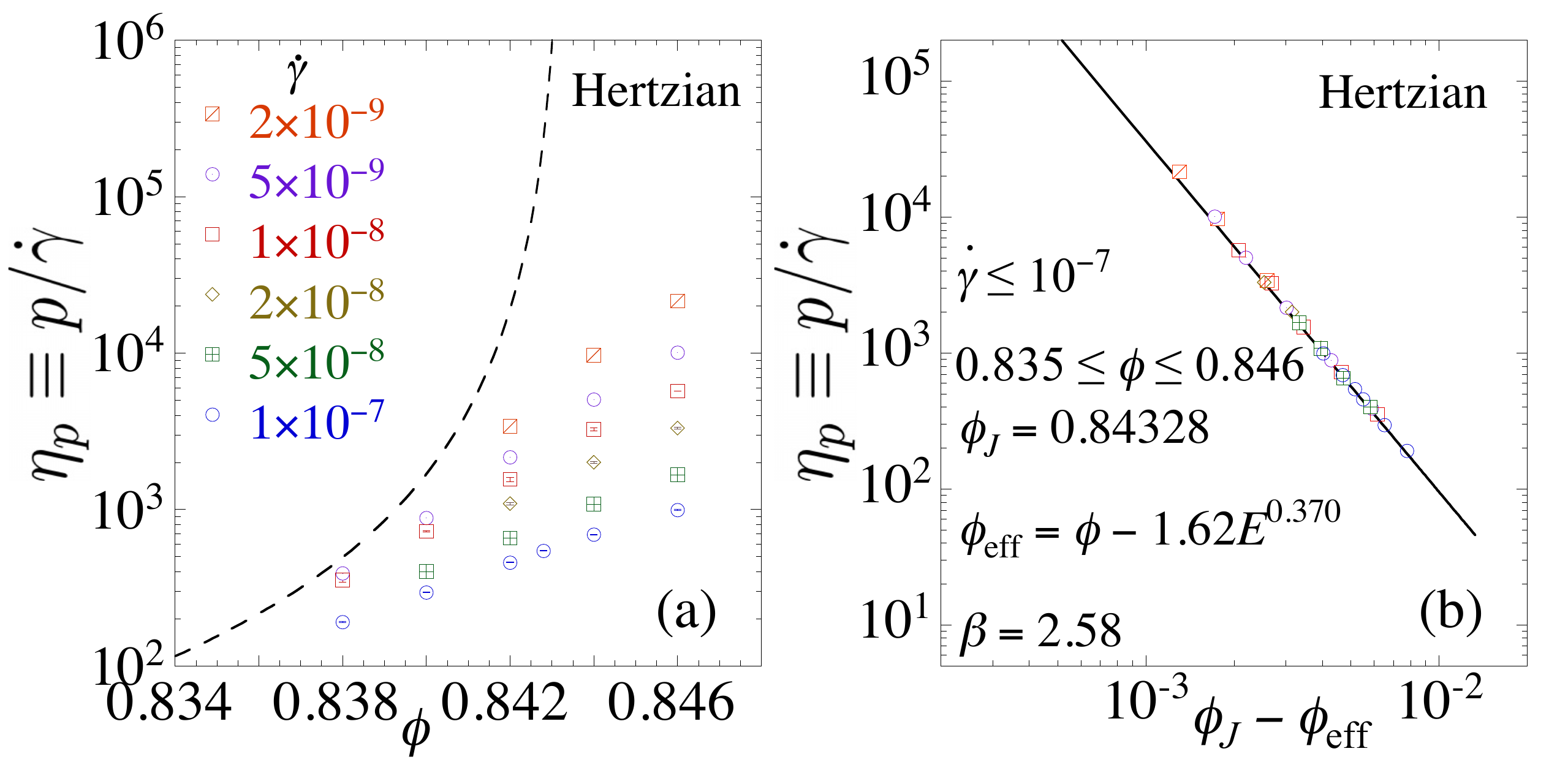}
\caption{(color online) (a) $\eta_p\equiv p/\dot\gamma$ vs $\phi$ for data above and below $\phi_J$ for Hertzian soft core particles. (b) Same data as in (a) but plotted vs $\phi_J-\phi_{\rm eff}$. Dashed line in (a) and solid line in (b) is the fit to the model of Eqs.~(\ref{e-etap}) and (\ref{e-phieff4}), using the same values of $A$, $\phi_J$ and $\beta$ as found for the harmonic interaction model. Symbols in (b) correspond to the legend given in (a).
}
\label{f3}
\end{center}
\end{figure}

Fig.~\ref{f3} shows a similar analysis for the Hertzian interaction. Outside the linear rheology region, the Hertzian $\eta_p$  is smaller than for the harmonic case due to the softer repulsion of the Hertzian cores.  Consequently, our Hertzian data generally lies further from the asymptotic $\dot\gamma\to 0$ hard-core limit  and thus is poorer at determining the critical behavior.
However, since the parameters $A$, $\phi_J$ and $\beta$ defining $\eta_p^{\rm hd}$ in Eq.~(\ref{e-etap}) are characteristic of the hard-core limit, only the parameters $c$ and $y_E$ defining $\phi_{\rm eff}$ in Eq.~(\ref{e-phieff4}) should vary as the soft-core interaction is changed. We therefore fix $A$, $\phi_J$ and $\beta$ to the values found from the harmonic data, and allow only $c$ and $y_E$ to vary.  The fit, shown in Fig.~\ref{f3}b,  is excellent and gives $y_E=2.70\pm 0.04$.  We find $y_E^{\rm Hertzian}/y_E^{\rm harmonic}=1.24\pm 0.05$, in good agreement with the ratio $\alpha^{\rm Hertzian}/\alpha^{\rm harmonic}=1.25$.  Since $E$ is related to the average particle overlap $\delta$ by $E\sim \delta^\alpha$, this observation suggests $\delta\sim(\phi-\phi_J)^{1.08}$, common to all soft-core interactions.

We now discuss the implications of our $\phi_{\rm eff}$ mapping for the Herschel-Bulkley rheology above $\phi_J$.  
For observables $X$ such as $p$, $\sigma$ or $E$,  the HB form for small $\dot\gamma$ is,
\begin{equation}
X(\phi,\dot\gamma)=X_0(\phi)+C_{X}\dot\gamma^{b_{X}}
\label{e-HB}
\end{equation}
where $b_X$ is the HB exponent, and $X_0(\phi)\equiv X(\phi,\dot\gamma\to 0)$.  First we review some conclusions that follow from a general critical scaling ansatz \cite{OlssonTeitelPRL,OlssonTeitelPRE}, independent of our $\phi_{\rm eff}$ mapping.  $X$ is expected to have the scaling form,
\begin{equation}
X(\phi,\dot\gamma)=|\delta\phi|^{y_X}{\cal X}_\pm\left(\dfrac{\dot\gamma}{|\delta\phi|^{z\nu}}\right)\enspace,\quad \delta\phi\equiv \phi-\phi_J\enspace,
\label{e-scal}
\end{equation}
where $z\nu=\beta+y_p$ \cite{OTnote} and ${\cal X}_\pm$ are the scaling functions for $\delta\phi \gtrless0$.
As $\phi\to\phi_J$, and the argument of the scaling function diverges, the dependence of $X$ on $\delta\phi$ should drop out, thus requiring, 
\begin{equation}
{\cal X}_\pm(x\to\infty)\sim x^{y_X/z\nu}\enspace,
\label{e-bigx}
\end{equation}
so that exactly at $\phi_J$ we have the non-linear rheology,
\begin{equation}
X(\phi_J,\dot\gamma)\sim \dot\gamma^{q_X}\enspace,\quad{\rm with}\quad q_X\equiv \dfrac{y_X}{z\nu}=\dfrac{y_X}{y_p+\beta}\enspace.
\label{e-q}
\end{equation}
Unlike $\beta$, we see that the rheology exponent $q_X$ at $\phi_J$ {\it does} depend on the particular soft-core interaction, via the exponents $y_p$ and $y_X$.

For $\phi>\phi_J$, as $\dot\gamma\to 0$, Eq.~(\ref{e-HB}) requires the scaling function ${\cal X}_+$ to have the form,
\begin{equation}
{\cal X}_+(x\to 0)= c_X+c_X^\prime x^{b_X}\enspace,
\label{e-smallx}
\end{equation}
with $c_X, c_X^\prime$ constants, 
so that we recover Eq.~(\ref{e-HB}) with
\begin{equation}
X_0=c_X\delta\phi^{y_X}\enspace,\quad{\rm and}\quad C_X=c_X^\prime\delta\phi^{y_X-b_Xz\nu}\enspace.
\end{equation}
Thus the coefficient $C_X$ of the HB law of Eq.~(\ref{e-HB}) must have a particular  scaling dependence on $\phi$ as $\phi\to\phi_J$.

We now return to our $\phi_{\rm eff}$ model and consider the pressure.  From the definition of $\eta_p$ and Eq.~(\ref{e-etap}) we can write,
 \begin{equation}
 p(\phi,\dot\gamma)=\dfrac{\dot\gamma A}{ (\phi_J-\phi_{\rm eff})^{\beta}}
 =\dfrac{\dot\gamma A}{ (\phi_J-\phi+h(E))^{\beta}}
 \label{e-xx}
 \end{equation}
Substituting in Eq.~(\ref{e-HB}) for $E$, and expanding $h(E)$ to first order for small $\dot\gamma$, we get,
\begin{equation} 
p(\phi,\dot\gamma)
= \dfrac{\dot\gamma A}{[h^\prime(E_0)C_E\dot\gamma^{b_E}]^\beta} \label{e-p}
\end{equation}
where we used $h(E_0)=cE_0^{1/y_E}=(\phi-\phi_J)$ to cancel out the leading term in the expansion of $h(E)$.  As $\dot\gamma\to 0$ above $\phi_J$, $p \to p_0$ is {\it finite}.  We thus conclude from Eq.~(\ref{e-p}) that we  must have, $b_E=1/\beta$.  

To determine the HB exponent for pressure, we just extend the expansion in Eq.~(\ref{e-p}) to next order,
\begin{eqnarray}
p&=&\dfrac{\dot\gamma A}{ [h^\prime(E_0)C_E\dot\gamma^{b_E}+
\frac{1}{2}h^{\prime\prime}(E_0) C_E^2\dot\gamma^{2b_E}]^\beta} \\ 
&\approx&p_0\left[1-\dfrac{\beta h^{\prime\prime}(E_0) C_E}{2h^\prime(E_0)}\dot\gamma^{b_E}\right]
\end{eqnarray}
Comparing to Eq.~(\ref{e-HB}) for $p$ we conclude that $b_p=b_E=1/\beta$.  
Similar results hold for the yield stress $\sigma$.  We thus conclude that the HB exponents are all equal to $b=1/\beta\approx 0.38\pm 0.02$, and by our earlier arguments, they are all independent of the particular soft-core interaction.
We note that a similar value, $b\approx 0.36$, was recently reported in experiments on sheared foams \cite{Mobius}.  

\begin{figure}
\begin{center}
\includegraphics[width=3.5in]{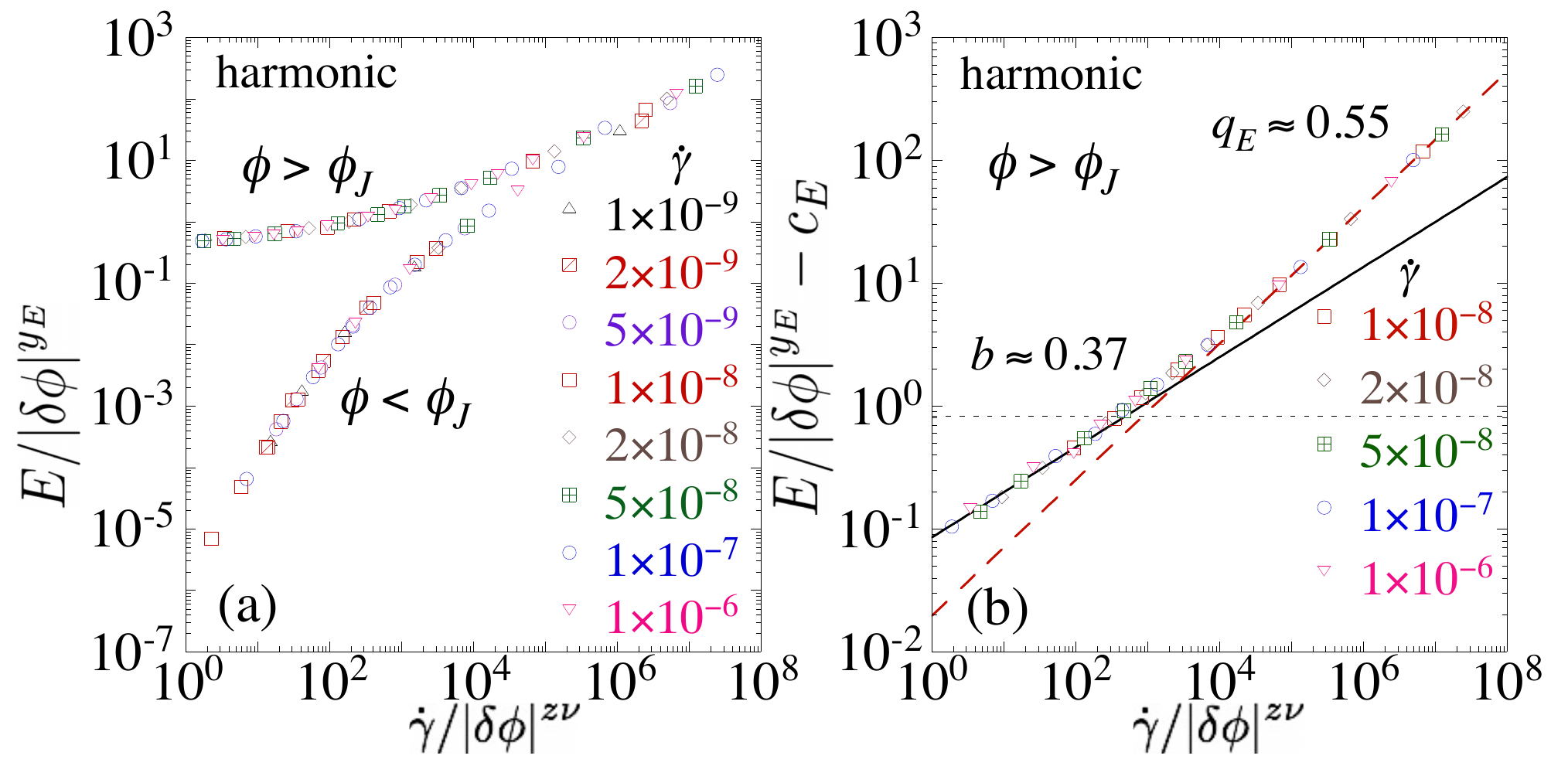}
\caption{(color online) (a) Scaling collapse of energy as in Eq.~(\ref{e-scal}), using values $\phi_J$, $y_E$, and $z\nu=\beta+y_p$ obtained from our fit to the $\phi_{\rm eff}$ model.  Two branches of the scaling function correspond to $\phi$ above and below $\phi_J$. (b) Plot of $E/|\delta\phi|^{y_E}-c_E$ vs $x\equiv\dot\gamma/|\delta\phi|^{z\nu}$ for $\phi>\phi_J$.  Dashed line is a fit to the large $x$ data, giving a power-law behavior with exponent $q_E\approx 0.55$; solid line is a fit to the small $x$ data, giving a power-law behavior with exponent $b\approx 0.37$.  Horizontal short dashed line is the value $c_E$; data below this line satisfy $(E-E_0)/E_0<1$.
}
\label{f4}
\end{center}
\end{figure}

We next numerically check our prediction for $b$. In Fig.~\ref{f4}a we show the scaling collapse of energy $E$ according to Eq.~(\ref{e-scal}) for the harmonic case. Using the parameters found from our $\phi_{\rm eff}$ fit, we see an excellent data collapse.  In Fig.~\ref{f4}b we plot $\tilde{\cal E}_+\equiv E/|\delta\phi|^{y_E}-c_E$ vs $x\equiv\dot\gamma/|\delta\phi|^{z\nu}$ for $\phi>\phi_J$,
where $c_E$ of Eq.~(\ref{e-smallx}) is obtained from $c$ of Eq.~(\ref{e-phieff4}) via $c_E=1/c^{y_E}$.  
From Eq.~(\ref{e-smallx}) we expect $\tilde{\cal E}_+\sim x^b$ at small $x$, while from Eq.~(\ref{e-bigx}) we expect $\tilde{\cal E}_+\sim x^{q_E}$ at large $x$.  
We consider $E$ rather than $p$ since there is a greater difference between the exponents $b$ and $q_E$ than  between $b$ and $q_p$.
Fitting the data of Fig.~\ref{f4}b separately at small and large $x$ we find power-law behaviors with $b\approx 0.37$ and $q_E\approx 0.55$, respectively,  
in reasonable agreement with the values expected from our $\phi_{\rm eff}$ analysis, $b=1/\beta=0.38$ and $q_E=y_E/(\beta+y_p)=0.59$.  The horizontal dashed line in Fig.~\ref{f4}b locates the value $c_E$ on the vertical axis. Data below this line satisfy the condition $(E-E_0)/E_0 < 1$.  We see that this condition locates the crossover from small to large $x$ behavior, which occurs near $x\approx 10^3$, or when $\dot\gamma\approx 10^3 \delta\phi^{z\nu}$.

Although we go to smaller values of $\dot\gamma$ than are typically used in other works, the closest our data for $\phi>\phi_J$ approaches the yield stress line is $(p-p_0)/p_0\gtrsim 0.18$.  One can always question whether this is close enough to give the true asymptotic critical behavior, or whether
rheological behavior might change at even smaller $\dot\gamma$.  We leave further investigation of this point to future work.  Here we note that experimental fits to the HB form usually involve data extending well above this limit down to values that typical do not go below $\approx 0.1$ \cite{Manneville}.  Thus, even if our $\phi_{\rm eff}$ model ultimately breaks down closer to the yield stress line, our results remain of considerable relevance for understanding the experimentally determined value of the HB exponent in numerous physical systems.

Our analysis has been for a model with dissipation to an external reservoir, yielding a Newtonian (linear) rheology below $\phi_J$. In athermal granular systems, with collisional dissipation and inertial effects, one expects Bagnold scaling \cite{Bagnold}, $\sigma, p\sim \dot\gamma^2$. In this case  the Bagnold coefficient scales as $B_p\equiv p/\dot\gamma^2\sim (\phi_J-\phi)^{-\beta^\prime}$ as jamming is approached \cite{Bagnold2}.  If a similar $\phi_{\rm eff}$ model holds, one can repeat all the steps of our above argument to arrive at the HB exponent for this case, $b=2/\beta^\prime$, while exactly at $\phi_J$ we have $q_X=2 y_X/(y_p+\beta^\prime)$. From Ref.~\cite{Bagnold2} we expect $\beta^\prime\approx 4$, giving $b\approx 0.5$.  We note that the exponent $b\approx 0.5$ was observed in recent molecular dynamic simulations of a 2D athermal Lenard-Jones glass 
\cite{Caroli}.

To conclude, by mapping soft-core particles at general $(\phi,\dot\gamma)$ to hard core particles at $(\phi_{\rm eff},\dot\gamma\to 0)$, we map the nonlinear rheology as $\dot\gamma\to 0$ above jamming to the linear rheology as $\dot\gamma\to 0$ below jamming, resulting in a relation between the HB exponent $b$ and the viscosity exponent $\beta$.  When comparing our results to experiments, however, several points must be kept in mind: (i) The HB exponent $b$ found here characterizes the rheology only for sufficiently small $\dot\gamma$.  Near $\phi_J$, as $\dot\gamma$ increases, one crosses into a region characterized by the exponent $q$ of Eq.~(\ref{e-q}) (see Fig.~\ref{f4}).  For systems with significant collisional dissipation and inertial effects, a crossover from Newtonian to Bagnold rheology is also possible \cite{Fall,Lemaitre2,Andreotti}.  Fitting the HB form to data that spans such crossover regions will therefore result in an {\it effective} exponent $b$ different from that reported here. (ii) The numerical value of $b$ we report here results from the simple Durian ``mean-field" model of dissipation, Eq.~(\ref{eDurian}).  Different models for viscous dissipation may yield different values for the exponent $\beta$, and hence for $b$ \cite{Andreotti,Tighe}. 

We  thank L. Berthier, D. Durian, T. K. Haxton, C. Heussinger, A. J. Liu, C. E. Maloney, S. Manneville and B. P. Tighe for helpful discussions.  
ST wishes to thank the Aspen Center for Physics, where some of these discussions were initiated.
This work was supported by Department of Energy Grant No. DE-FG02-06ER46298 and Swedish Research Council Grant No. 2010-3725.
The simulations were performed on resources
provided by the Swedish National Infrastructure for Computing (SNIC)
at PDC and HPC2N.  


\begin{thebibliography}{99}
\bibitem{footnote}We have verified that including corrections to scaling does not 
significantly alter the values of our fitted parameters.  The error estimates cited for our fitted parameters reflect the variations we find when including corrections to scaling.
\bibitem{OTnote} The relation $\beta=z\nu -y_p$ follows from applying the general Eq.~(\ref{e-scal}) to the observable $p$ and noting that in the linear rheology region of overdamped particles below $\phi_J$, the scaling function must be linear in its argument at sufficiently small $\dot\gamma$.  Hence we find $p/\dot\gamma\sim |\delta\phi|^{-(z\nu - y_p)}$.
\end{thebibliography}
\end{document}